\documentclass[conference]{IEEEtran}
\IEEEoverridecommandlockouts
\usepackage{cite}
\usepackage{amsmath,amssymb,amsfonts}
\usepackage{algorithmic}
\usepackage{graphicx}
\usepackage{textcomp}
\usepackage{xcolor}
\def\BibTeX{{\rm B\kern-.05em{\sc i\kern-.025em b}\kern-.08em
    T\kern-.1667em\lower.7ex\hbox{E}\kern-.125emX}}
\begin{document}

\title{Analytical Derivation of Quantization Error in Threshold Level Quantizers Using Bipolar PFM\\
\thanks{Research funded by grant PDC2023-145850-100 of the Spanish Ministry of Science, Innovation and Universities}
}
\author{\IEEEauthorblockN{1\textsuperscript{st} Ricardo Carrero}
\IEEEauthorblockA{\textit{Electronic Technology Dep.} \\
\textit{Carlos III University}\\
Madrid, Spain \\
rcarrero@pa.uc3m.es}
\and
\IEEEauthorblockN{2\textsuperscript{nd} Ruben Garvi}
\IEEEauthorblockA{\textit{Electronic Technology Dep.} \\
\textit{Carlos III University}\\
Madrid, Spain \\
rgarvi@ing.uc3m.es}
\and
\IEEEauthorblockN{3\textsuperscript{rd} Luis Hernandez}
\IEEEauthorblockA{\textit{Electronic Technology Dep.} \\
\textit{Carlos III University}\\
Madrid, Spain \\
luish@ing.uc3m.es}
}

\maketitle

\begin{abstract}
Uniform quantization is a topic that has been extensively studied. However and although an analytical description of quantization noise has been proposed, most descriptions of the spectral properties of quantization error resort to statistical descriptions. In this paper, we show that the spectrum of a quantized signal can be expressed using pulse frequency modulation. We first establish the equivalence of a uniform quantizer with a system based on the bipolar pulse frequency modulation and we define afterwards the Fourier transform of the quantized signal using pulse frequency modulation properties. This model brings a more intuitive understanding of the spectral structure of quantization noise and complements prior research in the topic. The results of the paper can be directly applied to level crossing ADCs with zero-order-hold interpolators, giving an accurate estimation of their performance.
\end{abstract}


\section{Introduction}
Analysis of quantization noise in uniform quantizers is a topic addressed over the past decades in several works, as for instance \cite{Zierhofer09,Gray98,Gray90,Claasen_Jongepier, widrow61,srypad77}. In \cite{Zierhofer09} a closed form for the quantization noise of an arbitrary continuous-time signal is provided. However, most papers use a statistical approach to compute the spectrum of quantized and sampled signals. The Level crossing ADC is an architecture \cite{Tsividis10} suitable for low power edge processing  of time-sparse signals \cite{VanAasche24}. Although Signal to Quantization Noise Ratio (SQNR) of level crossing ADCs can be improved by sophisticated reconstruction algorithms \cite{Tsividis10}, zero-order-hold interpolation is frequently used due to its simplicity \cite{review_NUS}. In this case, level crossing ADCs can be modeled as unsampled uniform quantizers, which motivates the interest in its analytical description. 

Integral Pulse Frequency Modulation (PFM) has been linked to sigma-delta modulation \cite{Medina_CASM}. In this paper, we investigate the possibility of a similar equivalence scheme between PFM and uniform quantization. We analyze an open loop uniform quantizer, showing that threshold quantization resorts to a pulse frequency modulation as well, although referred to the derivative of the input signal. With this PFM model we will calculate a series expansion of the output of a uniform quantizer for a sinusoidal input. This series shows that the quantization error spectrum is composed of an infinite number of input harmonics organized in modulation sidebands. The amplitudes of the input harmonics in the modulation sidebands are described by a series of Bessel functions centered around DC. As a fundamental difference to \cite{Medina_CASM}, use of bipolar pulse frequency modulation is compelling in the analysis of a uniform quantizer and is the reason why modulation sidebands appear all centered at DC. The paper first establishes a PFM-based model of a uniform quantizer, proving the need for a bipolar PFM for the model to be correct. Afterwards, the mathematical definition of bipolar PFM is laid out, and the existing mathematical description of unipolar PFM is extended to bipolar PFM. Finally we validate our model confronting simulation cases with theoretical predictions. The results of this paper allow to model accurately the operation of a level crossing ADC with zero-order-hold interpolator.

\section{PFM equivalent of a uniform quantizer}

As pointed out in the introduction, we have to distinguish between threshold level quantization and sampling. We are going to refer here to threshold level quantization but without considering a sampler afterwards, as a difference to \cite{Gray90}. As start point, we will show that the integral of a PFM encoded signal is identical to the quantized integral of said signal. 

\subsection{PFM Encoded Integrator}

\begin{figure}[t]
\centering
{\includegraphics [width=0.7\columnwidth]{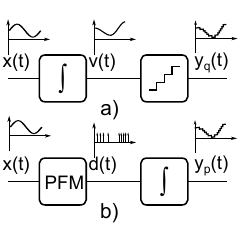}}
\caption{a) Integrator followed by uniform quantizer. b) PFM equivalent of a) } 
\label{fig:int_PFM_equiv}
\end{figure} 

Fig. \ref{fig:int_PFM_equiv}.a shows an integrator that integrates signal $x(t)>0$. The output, $v(t)$, is then quantized with a uniform quantizer with quantization step $\Delta$, producing signal $y_q(t)$. Fig. \ref{fig:int_PFM_equiv}.b, shows a PFM encoder that encodes the same signal $x(t)$ with PFM modulation \cite{PFMI16}. In \cite{baylySpectralAnalysisPulse1968, Medina_CASM} the typical system employed to implement a PFM modulator is described as a feedback loop involving integration and a threshold detector. The output of a PFM modulator consist of a train of Dirac delta impulses $d(t)$ at instants $t_k$ defined by the equation \cite{baylySpectralAnalysisPulse1968}:

\begin{eqnarray}
d(t)= \sum _{k=0} ^\infty \Delta \cdot \delta(t-t_k)  \nonumber  \\ 
t_k \, \mid \ \int_{t_{k-1}} ^{t_k} x(t)\,dt = \Delta 
\label{eq:pfm_recurrence}
\end{eqnarray}

We have a Dirac delta impulse in $d(t)$ every time the integral of $x(t)$ increases by an amount $\Delta$. The main property of PFM modulation is that the average value of the Dirac delta impulses in $d(t)$ follows the input signal, being a simple way to encode a low-pass continuous signal into a pulsed one. Observing the definition in \eqref{eq:pfm_recurrence}, it is clear that we have a delta impulse in $d(t)$ of Fig. \ref{fig:int_PFM_equiv}.b every time signal $v(t)$ in Fig. \ref{fig:int_PFM_equiv}.a crosses a threshold of the uniform quantizer. Now we integrate $d(t)$ which turns the summation of delta impulses into a summation of step functions $y_p(t)$:

\begin{eqnarray}
y_p(t)=\Delta \cdot \sum _{k=0} ^\infty \ u(t-t_k) 
\label{eq:quantized_integral_out}
\end{eqnarray}

where $u(t)$ represents the Heaviside step function. As can be seen, signals $y_p(t)$ and $y_q(t)$ are mathematically identical and hence, the systems of Fig. \ref{fig:int_PFM_equiv}.a and Fig. \ref{fig:int_PFM_equiv}.b can be considered equivalent. Note that this equivalence does not stem from the low pass signal approximation property of PFM modulation but from a mathematical identity. 

\subsection{Differential-Integral model of a threshold level quantizer}

\begin{figure}[t]
\centering
{\includegraphics [width=0.7\columnwidth]{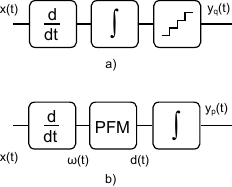}}
\caption{a) Uniform quantizer. b) Differential-Integral model.} 
\label{fig:differential_integral}
\end{figure}

The model of Fig. \ref{fig:int_PFM_equiv} shows the connection between quantization and PFM but requires an integration. To model a conventional uniform quantizer, we can complement the model using a differentiation function cascaded with the integration, as shown in Fig.\ref{fig:differential_integral}.a. The integration function in front of the quantizer would be compensated by the differentiation block. This setup, can be reproduced in the PFM equivalent by placing a differentiator in front of the PFM encoder, which now encodes signal $w(t)$, the derivative of $x(t)$. For simplicity, we will assume $x(t)$ to be zero mean in the foregoing. This configuration has however, some inconsistencies regarding the definition of the PFM modulator which need to be solved. Regardless of the input signal DC level, the DC content of signal $w(t)$ will always be zero due to the derivative. Hence, signal $w(t)$ will have positive and negative values depending of the slope of $x(t)$, but the definition in \eqref{eq:pfm_recurrence} requires the PFM input to be always positive. 
As a consequence, we need to modify the definition of the PFM signal to represent both positive and negative signals. Note that adding a DC level to $w(t)$ would not solve the problem as the threshold crossing points of $y_q(t)$ would no longer be coincident with those of $y_p(t)$. A feasible solution instead is to redefine the PFM modulation as follows:

\begin{eqnarray}
d(t)= \sum _{k=0} ^\infty P_k \cdot \delta(t-t_k)  \nonumber  \\ 
P_k = \int_{t_{k-1}} ^{t_k} x(t)\,dt \nonumber \\
t_k \, \mid \,  \left| P_k \right| = \Delta 
\label{eq:bpfm_recurrence}
\end{eqnarray}

where $\left| \cdot \right|$ represents the absolute value. This definition of the pulse frequency modulation will be referred as Bipolar PFM and retains the property of encoding the low pass contents of the input signal in the average value of the input signal. 

\begin{figure}[t]
\centering
{\includegraphics [width=0.85\columnwidth]{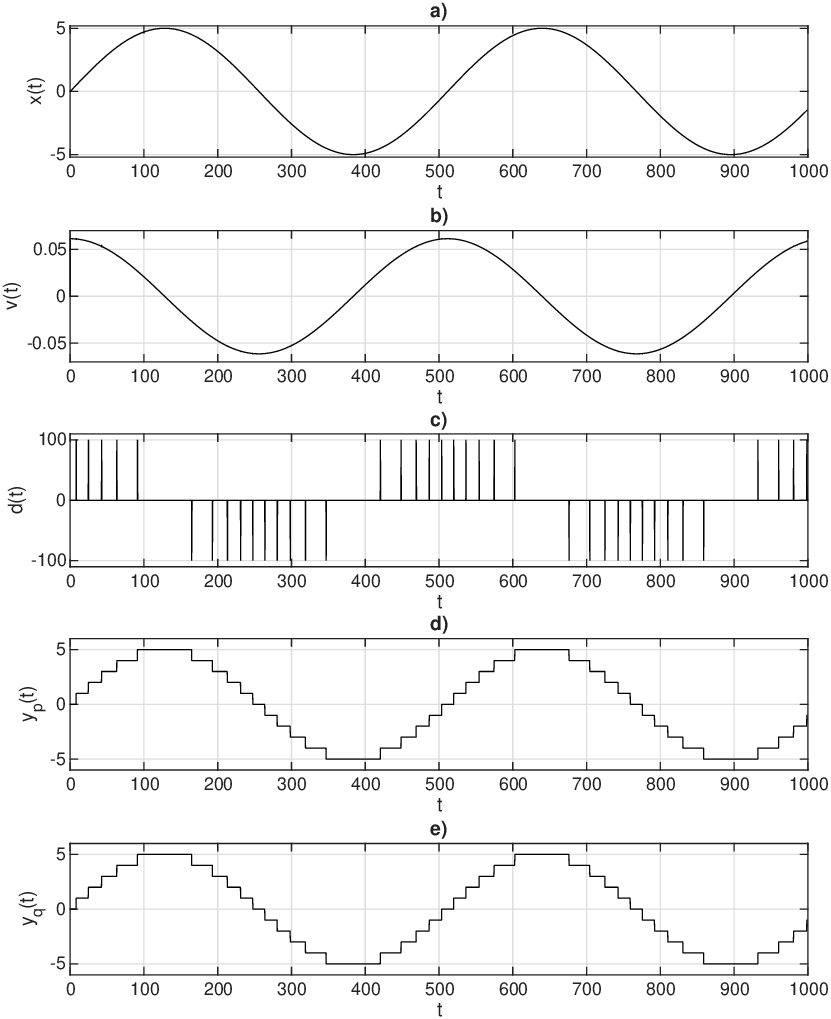}}
\caption{a) Input sinusoid. b) Input derivative. c) bipolar PFM. d) Signal $y_p(t)$. e) Signal $y_q(t)$} 
\label{fig:bpfm_demo}
\end{figure}

Fig. \ref{fig:bpfm_demo} shows a simulation of the system of Fig. \ref{fig:differential_integral} using an input sinusoid $x(t)$ with an amplitude A=5, frequency $f_x=2mHz$ (see Fig. \ref{fig:bpfm_demo}.a) and a quantizer with quantization step $\Delta=1$, which produces signal $y_q(t)$, plotted in Fig. \ref{fig:bpfm_demo}.e. After derivation, signal $v(t)$ is plotted in Fig. \ref{fig:bpfm_demo}.b. The bipolar PFM signal, $d(t)$, is represented in Fig. \ref{fig:bpfm_demo}.c, where we can see that delta impulses are coincident with the transitions between quantization levels of $y_q(t)$. Finally, after integration of $d(t)$, we obtain $y_p(t)$, plotted in Fig. \ref{fig:bpfm_demo}.d, which is equal to $y_q(t)$. 

Applying a DC input to the differential-integral model is possible but requires some further explanation. If we apply to a uniform quantizer a signal whose DC content is not zero, there will be also a quantized DC offset at the output. Given the input derivative in Fig. \ref{fig:differential_integral}.b one could think that no DC input can be encoded in our model. However, if we assume that the input is causal and there is a step at $t=0$ when the DC value is applied, there will be an input impulse at the derivative output which will set the proper initial value in the PFM modulator internal integrator translating in an initial integer number of delta pulses in $d(t)$ accounting for the right DC quantized value. 

\section{Spectral Analysis of PFM Quantizer Model}

The bipolar PFM quantizer model described in Fig. \ref{fig:differential_integral} does not represent more than just an alternative point of view of quantization. However, this representation allows to use the well known spectral analysis of PFM signals to study the spectral structure of quantization error \cite{Gray98,Claasen_Jongepier} in a more intuitive way, also providing analytical results. A series expansion for PFM signals expressed as Dirac delta pulses encoding a sinusoid can be found for instance in \cite{baylySpectralAnalysisPulse1968}. This expansion can be extended to an arbitrary number of sinusoids \cite{nakao} enabling a broader signal context. Let's assume that an input sinusoid $v(t)$ with DC value $v_m$, amplitude $B$ and frequency $f_x$, is applied to a PFM modulator as defined in \eqref{eq:pfm_recurrence}, we can write:

\begin{eqnarray}
v(t)=v_m+B\cdot \cos(2\pi f_xt)
\end{eqnarray}
Then, signal $d(t)$ can be expanded into a trigonometric series as follows:

\begin{eqnarray}
d(t)=f_0+  B\cdot \cos(2\pi f_xt) + m(t),\nonumber \\ \nonumber m(t)=2f_0 \cdot \Delta \cdot \sum_{q=1}^{\infty}\sum_{r=-\infty}^{\infty}J_r\left ( \frac{q \cdot B}{f_x \cdot \Delta} \right ) \\ \cdot \left ( 1+\frac{rf_x}{qf_0} \right )\cos(2\pi (qf_0+rf_x)t)
\label{eq:pfm_series}
\end{eqnarray}

\begin{figure}[t]
\centering
{\includegraphics [width=0.8\columnwidth]{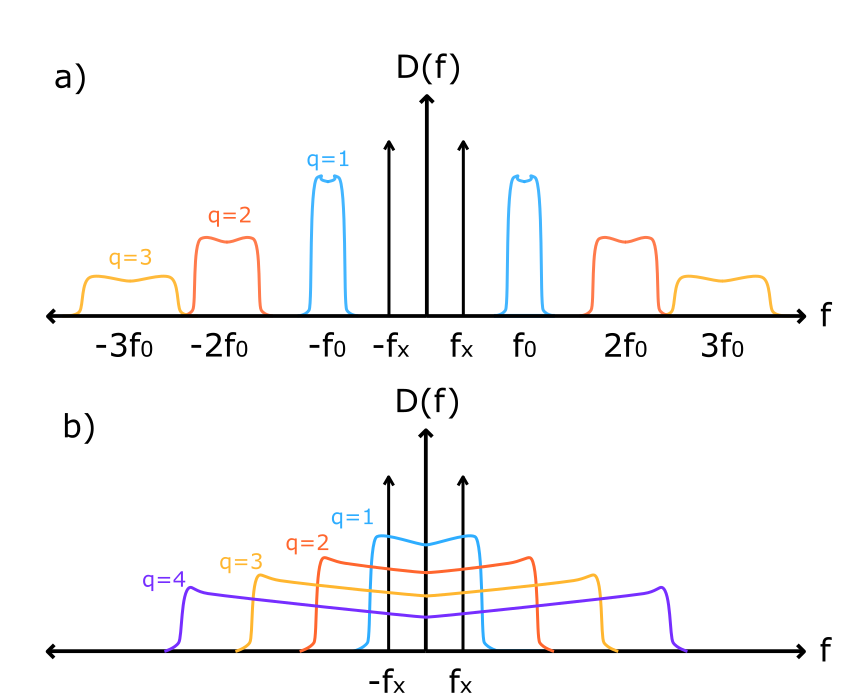}}
\caption{a) Single polarity PFM b) Bipolar PFM} 
\label{fig:pfm_vs_bpfm}
\end{figure} 

where $f_0$ corresponds to the rest frequency of the PFM modulator, which is the frequency produced when the DC value $v_m$ is applied, and $J_r(\cdot)$ are the Bessel functions of the first kind. This analysis shows that a PFM signal can be decomposed into two components, one representing the input signal itself and another composed of modulation sidebands $m(t)$. The modulation sidebands are groups of tones shifted by integer multiplies of $f_x$ around rest frequency $f_0$ and its harmonics. This situation is depicted in the example spectrum of Fig. \ref{fig:pfm_vs_bpfm}.a.

In the system of Fig. \ref{fig:differential_integral}, the signal has zero  mean by definition and then, $v_m$ and $f_0$ are zero. Note that in \cite{baylySpectralAnalysisPulse1968}, no restriction is placed in the values of $v_m$ and $f_0$ for the proof of \eqref{eq:pfm_series} to be valid, hence it applies to Bipolar PFM as well. Then, we can write:

\begin{eqnarray}
\nonumber x(t)= A \cdot sin(2 \pi f_xt) \\  
\nonumber v(t)= \frac{d(x(t))}{dt}= B \cdot cos (2 \pi f_xt)  \\  
\nonumber B=2 \pi f_x \cdot A    \\   
\nonumber d(t)=v(t)+m(t) \\
y_p(t)=x(t)+\int_{0} ^{t} m(\tau)\,d\tau
\label{eq:dif_int_relation}
\end{eqnarray}

 Applying \eqref{eq:dif_int_relation} to \eqref{eq:pfm_series} we obtain:

\begin{eqnarray}
d(t)=B \cdot \cos(2\pi f_xt) + m(t),\nonumber \\ m(t)=2\Delta \cdot \sum_{q=1}^{\infty}\sum_{r=-\infty}^{\infty}J_r\left ( \frac{q \cdot B}{f_x \cdot \Delta} \right ) \cdot \left (\frac{rf_x}{q} \right )\cos(2\pi rf_x t)
\label{eq:bpfm_series}
\end{eqnarray}

In \eqref{eq:bpfm_series} we can see that $m(t)$ is composed by harmonics of the input frequency $f_x$ only, weighted by the Bessel functions and indexes q and r. This situation is depicted in the example of Fig. \ref{fig:pfm_vs_bpfm}.b. The quantized signal can be obtained by integration. We can decompose the quantized signal $y_p(t)$ as the sum of the input signal $x(t)$ and the quantization error $e(t)$:

\begin{eqnarray}
y_p(t)= A \cdot \sin(2\pi f_xt) + e(t),\nonumber \\ e(t)=\Delta \cdot \sum_{q=1}^{\infty}\sum_{r=-\infty}^{\infty}J_r\left ( \frac{q \cdot 2\pi \cdot A}{ \Delta} \right ) \cdot \left (\frac{1}{\pi q} \right )\sin(2\pi rf_x t)
\label{eq:analytical_yp}
\end{eqnarray}

\begin{figure} 
\centering 
{\includegraphics [width=0.85\columnwidth]{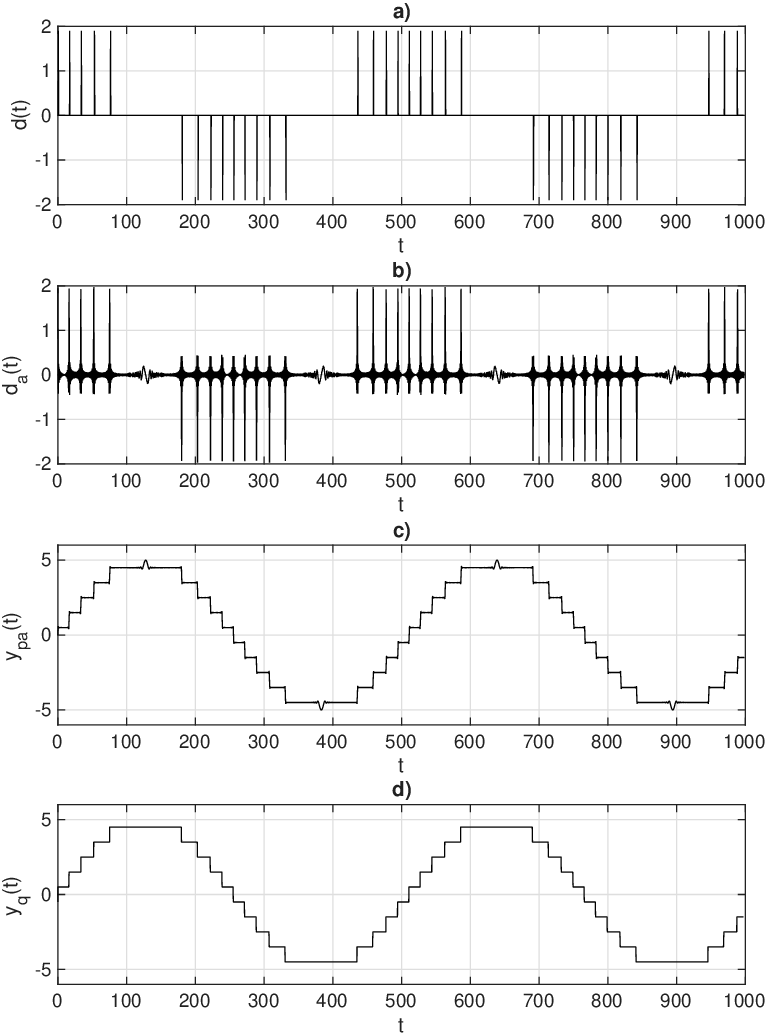}}
\caption{a) simulated PFM. b) Analytical PFM. c) Analytical quantized signal. d) Simulated quantized signal.} 
\label{fig:bpfm_test}
\end{figure} 

Since \eqref{eq:analytical_yp} consists of a sum of sine waves, we can analytically compute its Fourier transform, a result shown in \eqref{eq:analytical_yp_fourier}:

\begin{eqnarray}
Y_p(w)&=&A \cdot \pi \cdot \left( \delta(w+w_x) - \delta(w-w_x) \right)j + E(w), \nonumber \\[1ex]
E(w) &=& \Delta \cdot \sum_{q=1}^{\infty}\sum_{r=-\infty}^{\infty} 
J_r\!\left( \frac{q \cdot 2\pi \cdot A}{\Delta} \right) \cdot \frac{1}{q} \nonumber \\ 
&& \cdot \left( \delta(w+ r w_x) - \delta(w- r w_x) \right)j
\label{eq:analytical_yp_fourier}
\end{eqnarray}

Prior works mention that quantization noise can be expressed as frequency modulated sinusoids \cite{Zierhofer09} and the spectrum of uniformly sampled quantizers is linked to Bessel functions \cite{Gray90}. However, \eqref{eq:analytical_yp} applies to the continuous waveform produced by an unsampled threshold quantizer and hence, is valid for level crossing ADCs with zero-order-hold interpolation.

\section{Simulations}

To prove the calculations of the previous section, we have compared the time domain behavioral simulation of Fig. \ref{fig:bpfm_demo} with a numerical evaluation of \eqref{eq:bpfm_series} and \eqref{eq:analytical_yp}. Fig. \ref{fig:bpfm_test}.a reproduces the PFM signal of Fig. \ref{fig:bpfm_demo}.d for reference. Fig. \ref{fig:bpfm_test}.b shows the equivalent result computed analytically using \eqref{eq:bpfm_series}, which displays similar delta impulses. The simulation has been made truncating the infinite summations, using 1000 harmonics of $f_x$ (index r) and 50 modulation sidebands (index q). Fig. \ref{fig:bpfm_test}.c reproduces the same data for the series in \eqref{eq:analytical_yp} which can be contrasted with the original quantized signal, reproduced in Fig. \ref{fig:bpfm_test}.d. Due to the finite number of summation terms used in the series expansion, the Gibbs phenomena can be appreciated in Fig. \ref{fig:bpfm_test}.b as a ringing around the delta impulses. This effect was also previously identified in \cite{Zierhofer09}. As can be seen, the PFM analytical model accurately represents quantization using the well known PFM mathematical derivations. 

\begin{figure} 
\centering
{\includegraphics [width=0.8\columnwidth]{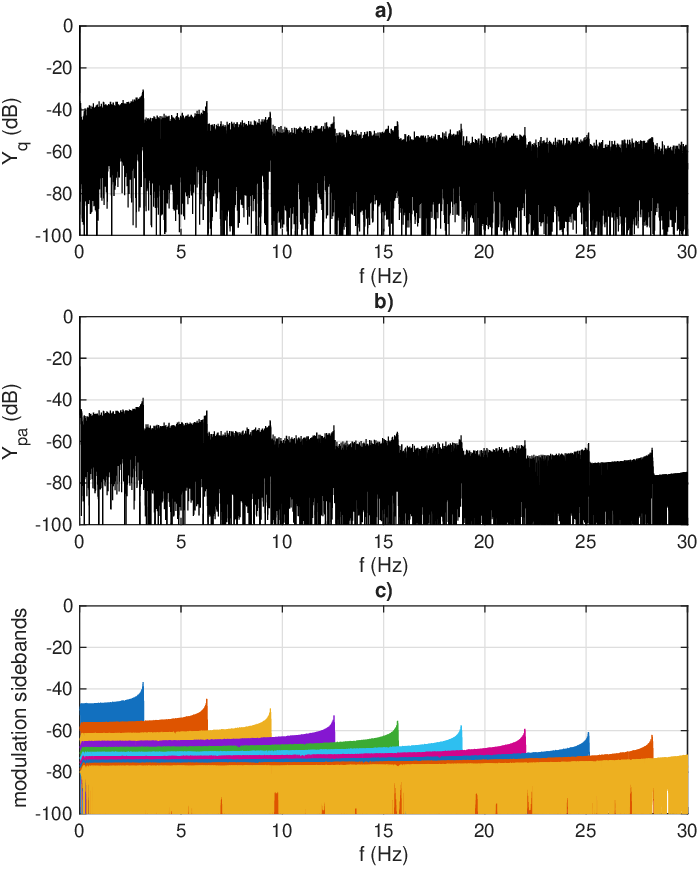}}
\caption{a) Simulated spectrum. b) Analytical spectrum. c) Sideband decomposition.}  
\label{fig:spectrum_eval}
\end{figure}

The series expansion of \eqref{eq:analytical_yp} can predict the spectral density of a quantized sinusoid, by representing the amplitude of each harmonic of $f_x$. We have simulated the system of Fig. \ref{fig:differential_integral} for a sine wave with amplitude A=512, representing the outcome of a 10 bit quantizer at full scale. Fig. \ref{fig:spectrum_eval}.a represents the FFT of the quantized sinusoid, suppressing the input tone. Fig. \ref{fig:spectrum_eval}.b represents the amplitudes of the harmonics of $f_x$, computed using \eqref{eq:analytical_yp} in dB. A similar spectrum can be observed, corresponding with the prediction of Fig. \ref{fig:pfm_vs_bpfm}.b. For further insight, Fig. \ref{fig:spectrum_eval}.c shows each modulation sideband spectrum, computed using (8) and displayed individually in dB, from q=1 (blue) up to q=10 (yellow).

\section{Conclusions}
Quantization is a well know process, but requires advanced mathematical tools to be described, resorting to statistical calculations in some cases. In this paper we have established the connection between quantization and Pulse Frequency Modulation. This novel point of view, provides an intuitive approach to estimate the spectral properties of quantized signals, by leveraging the well known PFM theory. In addition, we have derived an equation to describe analytically the quantization error of sinusoidal waveforms. This result can be directly applied to level crossing ADCs using zero-order-hold interpolation.
\vfill\null
\bibliography{./references}

\bibliographystyle{IEEEtran}


\end{document}